# A Proposed Algorithm for Minimum Vertex Cover Problem and its Testing

Gang Hu
Email: garyhu3@163.com

**Abstract**

The paper presents an algorithm for minimum vertex cover problem, which is an NP-Complete problem. The algorithm computes a minimum vertex cover of each input simple graph. Tested by the attached MATLAB programs, Stage 1 of the algorithm is applicable to, i.e., yields a proved minimum vertex cover for, about 99.99% of the tested 610,000 graphs of order 16 and 99.67% of the tested 1,200 graphs of order 32, and Stage 2 of the algorithm is applicable to all of the above tested graphs. All of the tested graphs are randomly generated graphs of random "edge density" or in other words, random probability of each edge. It is proved that Stage 1 and Stage 2 of the algorithm run in $O(n^{5+\log n})$ and $O(n^{3(5+\log n)/2})$ time respectively, where $n$ is the order of input graph. Because there is no theoretical proof yet that Stage 2 is applicable to all graphs, further stages of the algorithm are proposed, which are in a general form that is consistent with Stages 1 and 2.

## 1. INTRODUCTION

If the algorithm is to be classified as one algorithm design technique, then transform-and-conquer strategy maybe most suitable for it [1].

Suppose we need to find a minimum vertex cover of simple graph $G$, then the first part of the algorithm is to generate an auxiliary simple graph $H$, which satisfies the following four conditions:

*A.* $L(V(H)) \supseteq L(V(G))$, of which $L(X)$ denotes $X$'s label set, i.e., the set of $X$'s all vertex's labels;

*B.* Different vertices of $H$ may share a same label. For any edge of $H$, however, the two endpoints have different labels;

*C.* Of each component $P$ of $H$, all maximal-clique's vertex-sets have a same cardinal number, and thus this number is called "grade of component $P$" and denoted by $g(P)$;

*D.* For any nonempty subset $T$ of $L(V(G))$, there exists a maximal clique $Q$ in $H$ such that $L(V(Q)) = T$.

*Remark* 1.1. $H$ is generated by Steps 1 to 3 in Section 2, and it is not any other graph which satisfies the above conditions.

The remaining part of the algorithm is to find out a maximal clique of $H$ whose label set can be proved to be the label set of a minimum vertex cover of graph $G$.

Let $N[x] = N(x) \cup \{x\}$, of which $N(x)$ is the neighbor set of vertex $x$. Suppose a subgraph of $H$ which satisfies both of the following conditions.

**Condition 1.** For any vertex $v$ of the subgraph, there exists a vertex cover $C$ of $G$ such that $L(N[v]) = L(C)$.

**Condition 2.** For any edge $uv$ of the subgraph, there exists a vertex cover $C$ of $G$ such that $L(N[uv]) = L(C)$, where $N[uv] = N[u] \cap N[v]$.

Then for each component of $H$ in ascending order of its grade, which is defined in Condition $C$, find out a maximal subgraph $S$ of it which satisfies Conditions 1 and 2. Iterate this computation for the next component until $S$ is nonempty. Then find out a minimal-order subgraph $I$ of $S$ which contains a nonempty subgraph that satisfies Conditions 1 and 2. If $I$ is a clique, then it is proved by Claim 2.10 that $L(V(I))$ is the label set of a minimum vertex cover of graph $G$. If $I$ is not a clique, then go to Stage 2 of the algorithm, which is introduced in Section 5

In Section 3, Stage 1 of the algorithm is proved to run in $O(n^{5+\log n})$ time.

The algorithm was tested by MATLAB programs, and it was found that $I$ is not a clique for only 44 of the tested 711,200 graphs. Then by Claim 2.10, Stage 1 of the algorithm cannot yield a minimum vertex cover for only about 0.0062% of all tested graphs. The test results are detailed in Section 4.

For the graphs to which Stage 1 of the algorithm is not applicable, a stronger version of Condition 2 is introduced. With this change, the algorithm reaches Stage 2 for the graphs to which Stage 1 is not applicable, and it was found that, with $O(n^{3(5+\log n)/2})$ time, Stage 2 is applicable to, i.e., yields a minimum vertex cover of, each of the 44 tested graphs. Because Stage 2 is same as Stage 1 except that Condition 2 is replaced by a stronger version, so Stage 2 actually works for all of the 711,200 tested graphs.

Furthermore, if there exist graphs to which Stage 2 of the algorithm is not applicable, those graphs will go to further stages of the algorithm which are expressed in a general form that is consistent with Stages 1 and 2. Those stages are introduced in Section 6.

## 2. STEPS OF THE ALGORITHM

Firstly, graph $H$ is constructed by Steps 1 to 3.

**Step 1.** Suppose $L(V(G)) = \{1, 2, \cdots, n\}$, and $k = \lceil \log_2 n \rceil$, i.e., $k$ is the smallest integer which satisfies $n \leq 2^k$. Let $L(V(H)) = \{1, 2, 3, \cdots, 2^k\}$.

Then define $Z$, which is a family of sets, as follows.

**Definition 2.1** Let $L(V(H))$ be the "first" member of $Z$. Then partition $L(V(H))$ into two equal-sized disjoint sets $\{1, 2, \cdots, 2^{k-1}\}$ and $\{2^{k-1}+1, 2^{k-1}+2, \cdots, 2^k\}$, whose elements are consecutive numbers, and let the two sets become members of $Z$. For each of the two "new" members, if its cardinal number is larger than 1, then continue the partition and "member-assignment" process as described above. Keep on the process until the cardinal number of each "new" member is 1. ■

$Z$ can also be generated in a reversed way, which is shown in the attached MATLAB program.

For example, if $k = 3$, then $Z = \{\{1\}, \{2\}, \{3\}, \{4\}, \{5\}, \{6\}, \{7\}, \{8\}, \{1, 2\}, \{3, 4\}, \{5, 6\}, \{7, 8\}, \{1, 2, 3, 4\}, \{5, 6, 7, 8\}, \{1, 2, \cdots, 8\}\}$.

**Step 2.** $\forall\, l \in L(V(G))$, generate one and only one first-grade component of $H$ whose label is $l$, and denote this component by $P(\{l\})$.

*Remark* 2.2 The grade of $H$'s component is defined in Condition $C$ of Section 1, it is obvious that the set of all first-grade components of $H$ are actually the set of $H$'s all isolated vertices. ■

**Step 3.** Other components of $H$ are generated in sequence by their grades. For example, all of the second-grade components are to be constructed before the construction of any higher-grade component. Components of any grade larger than 1 are defined in a general form as follows. But firstly, another definition is needed.

**Definition 2.3.** Suppose $A$ is a member of $Z$, which is defined by Definition 2.1, and $b$ is a positive integer. Let $U = \cup P$, of which $P$ is $H$'s component which satisfies: $L(V(P)) \subseteq A$ and $g(P) = b$. Then we call $U$ a same-grade union of $A$ and $b$, and denote it by $U(A, b)$. (See Fig. 2.4.)

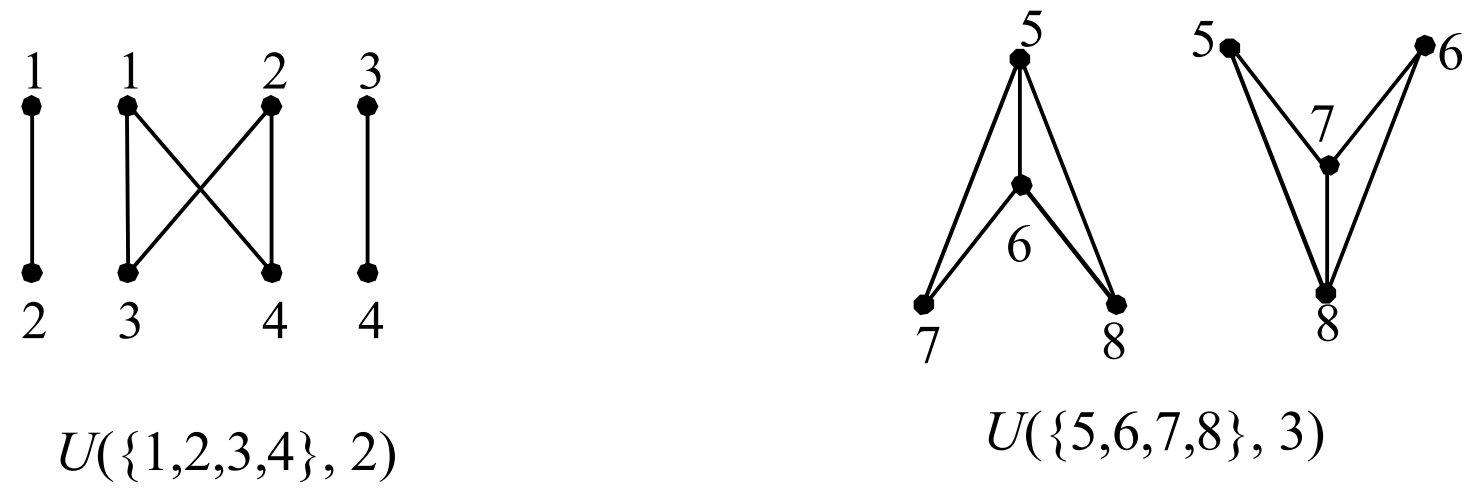

**Figure 2.4.** Examples of same-grade unions

Each component of $U(A, b)$, where $b$ is larger than 1, is defined as follows.

**Definition 2.5.** If and only if $P$ is a join of two same-grade unions $U(A_1, b_1)$ and $U(A_2, b_2)$, i.e., $P = U(A_1, b_1) \vee U(A_2, b_2)$, of which $A_1$, $A_2$, $b_1$ and $b_2$ satisfy the following two conditions:

2.5.1 $A_1 \cap A_2 = \emptyset$ and $A_1 \cup A_2 = A$, of which $A \in Z$;

2.5.2 $b_1$ and $b_2$ are positive integers, and $b_1 + b_2 = b$,

then let $P$ be one of $H$'s components denoted by $P(A_1, b_1, A_2, b_2)$ or $P(A, b_1, b_2)$, of which $b_1$ corresponds to $A$'s "first half" containing smaller numbers.

After and only after for the same $b$, $P$ of each $A_1$, $A_2$, $b_1$ and $b_2$ which satisfy conditions 2.5.1 and 2.5.2 is generated, we start to construct $H$'s components of grade $b$+1. When the highest-grade component of $H$, which is a $2^k$-clique, is generated, the construction of $H$ is complete, then we call $H$ a minimum-covering-computation graph. ■

Then we have

**Claim 2.6.** For any nonempty subset $T$ of $L(V(G))$ and any member $A$ of $Z$ which satisfies $T \subseteq A$, there exists a maximal clique $Q$ in $U(A, |T|)$ such that $L(V(Q)) = T$.

*Proof.* We use induction on $|T|$. The claim obviously holds when $|T| = 1$.

For the induction step, let $T$ be a subset of $L(V(G))$ with $|T| \geq 2$, then by Definition 2.1, we can suppose $A_m$ to be the only minimal member of $Z$ which includes $T$. Because $|T| \geq 2$, we can suppose that $A_m = A_1 \cup A_2$ of which $A_1$ and $A_2$ are also $Z$'s members and $A_1 \cap A_2 = \varnothing$, then let $T_1 = A_1 \cap T$ and $T_2 = A_2 \cap T$. Since $A_m$ is the minimal member of $Z$ which includes $T$, it is obvious that $T_1, T_2 \neq \varnothing$. By the induction hypothesis, there exists a maximal clique $Q_1$ in $U(A_1, |T_1|)$ such that $L(V(Q_1)) = T_1$, and there exists a maximal clique $Q_2$ in $U(A_2, |T_2|)$ such that $L(V(Q_2)) = T_2$. Then by Definition 2.5, there exists a maximal clique $Q$ in $P(A_1, |T_1|, A_2, |T_2|) = U(A_1, |T_1|) \vee U(A_2, |T_2|)$ such that $L(V(Q)) = L(V(Q_1) \cup V(Q_2)) = T_1 \cup T_2 = T$.

$P(A_1, |T_1|, A_2, |T_2|)$ is a component of $U(A_m, |T|)$. Because $A_m$ is the only minimal member of $Z$ which includes $T$, so for any member $A$ of $Z$ which includes $T$, $A$ includes $A_m$ and $P(A_1, |T_1|, A_2, |T_2|)$ is also a component of $U(A, |T|)$, thus there exists a maximal clique $Q$ in $U(A, |T|)$ such that $L(V(Q)) = T$. ■

We need a claim as follows to understand the next step.

**Claim 2.7.** For any subgraph $D$ of $H$, there is only one maximal subgraph of $D$ which satisfies Conditions 1 and 2 in Section 1.

*Proof.* Assume $S_1$ and $S_2$ are two maximal subgraphs of $D$ which satisfy Conditions 1 and 2, then $S_1 \nsubseteq S_2$ and $S_2 \nsubseteq S_1$. Let $S = S_1 \cup S_2$, then $S \supset S_1$ and $S \supset S_2$. However, it is obvious that $S$ also satisfies Conditions 1 and 2, which is contradictory to the assumption. ■

**Step 4.** For each component of $H$ in ascending order of its grade, which is defined in Condition $C$ of Section 1, by deleting a minimal subgraph of the component, get a maximal subgraph $S$ of it which satisfies Conditions 1 and 2. Suppose the grade of the current component is $b$. If $S$ is empty, iterate this computation for the next component, whose grade is also $b$ or otherwise $b$+1 when all components of grade $b$ have been computed. If $S$ is nonempty, record $S$ for the next step. ■

Then we have a claim as follows.

**Claim 2.8.** For the recorded $S$ of Step 4, the size of its maximum clique is no larger than the minimum size of graph $G$'s vertex cover.

*Proof.* Suppose $C$ is a minimum vertex cover of $G$, then by Claim 2.6, there exists a maximal clique $Q$ in $U(L(V(H)), |C|)$ such that $L(V(Q)) = L(C)$. Thus $Q$ satisfies Conditions 1 and 2, so there exists component of $U(L(V(H)), |C|)$ which has nonempty subgraph satisfying Conditions 1 and 2. Because each component $P$ of $H$ is computed in ascending order of its grade and by Claim 2.7 $S$ is the only maximal subgraph of $P$ which satisfies Conditions 1 and 2, so for the recorded nonempty $S$ of Step 4, the size of its maximum clique is no larger than $|C|$. ■

**Step 5.** For the recorded $S$ of Step 4, by deleting a maximal subset of $V(S)$, get a minimal-order subgraph $I$ of $S$ which contains a nonempty subgraph $SS$ that satisfies Conditions 1 and 2. Record this $I$ for the next step.

*Remark* 2.9. By Claim 2.7, we can tell whether $I$ contains a nonempty subgraph that satisfies Conditions 1 and 2 by finding out the only one maximal subgraph of $I$ which satisfies the two conditions. ■

Then for $I$ recorded by Step 5, we have a claim as follows.

**Claim 2.10.** If $I$ is a clique, then $L(V(I))$ is the label set of a minimum vertex cover of graph $G$.

*Proof.* By Step 5, $I$ contains a nonempty subgraph that satisfies Conditions 1 and 2, so by Condition 1 or 2, we can tell that there exists a vertex cover $C$ of $G$ such that $L(V(I)) = L(C)$.

Suppose $b$ is the size of a maximum clique of the recorded $S$ in Step 4. If $I$ is a clique, then $|V(I)| \leq b$ because $I$ is a subgraph of $S$. By Claim 2.8, $b$ and so $|V(I)|$ are no larger than the minimum size of graph $G$'s vertex cover. Thus $L(V(I)) = L(C)$ is the label set of a minimum vertex cover of $G$. ■

Thus we have

**Step 6.** Check whether the recorded $I$ of Step 5 is a clique. If $I$ is a clique, then output $L(V(I))$. If $I$ is not a clique, then go to the next stage of the algorithm.

*Remark* 2.11. Sections 5 and 6 explain "the next stage of the algorithm". ■

## 3. EFFICIENCY OF THE ALGORITHM

**Claim 3.1** For $L(V(H)) = \{1, 2, 3, \cdots, 2^k\} (k \geq 1)$,

$$|V(H)| = \prod_{i=1}^{k} (2^i + 2).$$

*Proof.* When $k = 1$, $H$ has two first-grade components and one second-grade component, and so $|V(H)| = 4 = (2^k+2)$.

We use induction on $k$ and assume the claim holds for any $k$ when $1 \leq k \leq h$.

Now for $k = h+1$, $L(V(H)) = \{1, 2, 3, \cdots, 2^{h+1}\}$. By Steps 1 to 3 of the algorithm, we can suppose two isomorphic minimum-covering-computation graph $H_1$ and $H_2$ which satisfy: $L(V(H_1)) = \{1, 2, 3, \cdots, 2^h\}$ and $L(V(H_2)) = \{2^h+1, 2^h+2, \cdots, 2^{h+1}\}$. Then by the induction hypothesis,

$$|V(H_1)| = \prod_{i=1}^{h} (2^i + 2).$$

Because $H_2$ is isomorphic to $H_1$, $|V(H_2)| = |V(H_1)|$. Suppose $P$ is a component of $H_1$ or $H_2$, then by Steps 1 to 3 of the algorithm, we can tell that $P$ is also a component of $H$. Thus we have

**Corollary 3.1.1.** Let $X = \cup P$, of which $P$ is $H$'s component and $P$ is also $H_1$'s or $H_2$'s component, then

$$|V(X)| = 2\prod_{i=1}^{h} (2^i + 2).$$

Now the other vertex-amount we need to count is of $H$'s components of which each is neither $H_1$'s nor $H_2$'s component. Suppose $P^*$ is such a kind of component, then by Definition 2.1 and 2.5, we can tell that $L(V(P^*)) \nsubseteq L(V(H_1))$, $L(V(P^*)) \nsubseteq L(V(H_2))$ and so $L(V(P^*)) = L(V(H))$. Thus by Definition 2.5,

$$P^* = U(A_1, b_1) \vee U(A_2, b_2) = U(L(V(H_1)), b_1) \vee U(L(V(H_2)), b_2).$$

In the above formula, for each $U(L(V(H_1)), b_1)$, $b_2$ of $U(L(V(H_2)), b_2)$ can be any integer from 1 to $|L(V(H_2))| = 2^h$, which means that each $U(L(V(H_1)), b_1)$ was copied for $2^h$ times to generate different components of $H$. Because

$$H_1 = \bigcup_{b_1=1}^{2^h} U\left(L\left(V\left(H_1\right)\right), b_1\right),$$

so $H_1$ was copied for $2^h$ times to construct $H$. And there is a same conclusion for $H_2$. Combine this conclusion with Corollary 3.1.1, then we have

$$\begin{aligned}|V(H)| &= |V(X)| + 2^h (|V(H_1)| + |V(H_2)|)\\ &= 2\prod_{i=1}^{h} (2^i+2) + 2^h\left(\prod_{i=1}^{h} (2^i+2) + \prod_{i=1}^{h} (2^i+2)\right)\\ &= \prod_{i=1}^{h+1} (2^i + 2).\end{aligned}$$

Thus the induction succeeds and the claim is proved. ■

**Claim 3.2.** Suppose graph $G$ of order $n$ is the input and $T_1(n)$ is the running time of Stage 1 of the algorithm, then $T_1(n) \leq O(n^{5+\log n})$.

*Proof.* By Step 1, $k$ is the smallest integer which satisfies $n \leq 2^k$, so $2^{k-1} < n$. Then by Claim 3.1,

$$|V(H)| = \prod_{i=1}^{k} (2^i + 2) \leq \prod_{i=1}^{k} (2^{i+1}) = 2^{k(k+3)/2} < (2\text{n})^{(log_2 n+4)/2}.$$

Thus the asymptotic upper bounds for $|V(H)|$ and $|E(H)|$ can be $n^{(5+\log n)/2}$ and $n^{5+\log n}$, respectively, so the construction of $H$ runs in $O(n^{5+\log n})$ time.

Then in Step 4, the running time for computing subgraphs of $H$'s components by Conditions 1 and 2 is still no more than $O(n^{5+\log n})$, and this is also the case for Step 5. Therefore Stage 1 of the algorithm takes no more than $O(n^{5+\log n})$ time. ■

## 4. TESTING OF THE ALGORITHM

The algorithm was tested by the two attached programs, "Generating_k_and_H" and "Testing_G", on MATLAB (R2016a, trial use). The first file generates and saves $k$ and $H$ for the second file. The second file generates random graphs of random "edge density", i.e., random probability of each edge, for testing of the algorithm. "Edge density" in the program approximately equals to the ratio of $|E(G)|$ to $|E(K)|$, of which $K$ is a $2^k$-order complete graph, and it is generated from a normal distribution with mean 0.5 and standard deviation 0.2,0.3,0.4,0.5 or 0.6 for the testing. The reasons for the introduction of this distribution are that there exist the most graphs, no matter labelled or unlabeled, when "edge density" equals to 0.5, and there exists a same amount of graphs of "edge density" being $d$ as that of "edge density" being $1 - d$.

An ordinary personal computer is used for the testing. Testing parameters and results are listed as follows.

**Table 4.1.** Testing parameters and results

| $k$, $\|V(G)\|$ | Number of tested graphs | Standard deviation of "edge density" | Approximate running time (for Testing_G.m) | No. of graphs to which Stage 1 is not applicable | No. of graphs to which Stage 2 is not applicable |
|---|---|---|---|---|---|
| 3, 8 | 100,000 | 0.3 | 16 minutes | 0 | 0 |
| 4, 16 | 50,000 | 0.2 | 10 hours | 2 | 0 |
| 4, 16 | 220,000 | 0.3 | 43 hours | 19 | 0 |
| 4, 16 | 150,000 | 0.4 | 29 hours | 8 | 0 |
| 4, 16 | 130,000 | 0.5 | 25 hours | 9 | 0 |
| 4, 16 | 60,000 | 0.6 | 12 hours | 2 | 0 |
| 5, 32 | 400 | 0.3 | 58 hours | 1 | 0 |
| 5, 32 | 400 | 0.4 | 58 hours | 2 | 0 |
| 5, 32 | 400 | 0.5 | 58 hours | 1 | 0 |

*Remark* 4.2. Stage 2 is explained in Section 5.

*Remark* 4.3. When $k = 5$, the running time for the computer used for testing is so long that it is impossible to test a large number of graphs in a short period. There is no test data for $k > 5$ for the same reason.

**Table 4.4.** Recorded "edge densities" of graphs to which Stage 1 is not applicable

| $k$, $\|V(G)\|$ | Recorded "edge densities" of graphs to which Stage 1 is not applicable |
|---|---|
| 4, 16 | 0.2410, 0.2599, 0.2973, 0.3170, 0.3173, 0.3204, 0.3323, 0.3359, 0.3533, 0.3551, 0.3642, 0.3773, 0.3774, 0.3942, 0.4089, 0.4232, 0.4436, 0.4475, 0.4495, 0.4500, 0.4509, 0.4539, 0.4821, 0.4919, 0.4938, 0.5032, 0.5318, 0.5337, 0.5343, 0.5356, 0.6200, 0.6476 |
| 5, 32 | 0.1479, 0.1889, 0.2127, 0.2868 |

By Table 4.1, when $k = 3$, Stage 1 is applicable to 100% of tested random graphs (i.e. Step 5 of Stage 1 yields a clique and, by Claim 2.10, Step 6 yields the label set of a minimum vertex cover of each tested graphs for $k = 3$); when $k = 4$, Stage 1 is applicable to about 99.99% of tested random graphs; and when $k = 5$, Stage 1 is applicable to about 99.67% of tested random graphs.

By Table 4.4, the recorded "edge densities" of tested graphs to which Stage 1 is not applicable scatter.

Except for $k = 3$, the numbers of tested graphs are much less than the numbers of unlabeled graphs of respective orders. However, because of the large number of tested graphs, and because the possibility of duplicated graphs to which the algorithm is applicable is same as that of duplicated graphs to which the algorithm is not applicable, it is reasonable to believe that the "applicability ratios" of tested graphs are close to the "applicability ratios" of all unlabeled (or labelled) graphs of respective orders.

## 5. STAGE 2 OF THE ALGORITHM

For those graphs to which Stage 1 of the algorithm is not applicable, a stronger version of Condition 2 is introduced as follows.

**Condition $2^+$.** For any edge $uv$ of the subgraph, there exists a vertex cover $C$ of $G$ such that $L(N[uv]) = L(C)$, where

$N[uv] = \{w \in N[u] \cap N[v]$ : There exists a vertex cover $C^*$ of $G$ such that $L(N[u] \cap N[v] \cap N[w]) = L(C^*)\}$. ■

With this change, the algorithm goes to Stage 2 for those graphs to which Stage 1 is not applicable. Steps 1 to 3 are for construction of $H$ and apparently they do not need to run again. Besides, for those components of $H$ which have been computed in Step 4 of Stage 1 and concluded that they do not have nonempty subgraph satisfying Conditions 1 and 2, they do not need to be computed again in Stage 2 because Condition $2^+$ is stronger than Condition 2. Thus Stage 2 starts at Step 4 for the component of $H$ which has nonempty subgraph satisfying Conditions 1 and 2, and the following computation is same as Stage 1 except that Condition 2 is substituted by Condition $2^+$.

Stage 2 of the algorithm has been implemented in the attached MATLAB program. It was found that Stage 2 yielded cliques for all of the 44 tested graphs for which Stage 1 did not. Because Claims 2.8 and 2.10 still hold for Stage 2, so Stage 2 is applicable to all of the 44 tested graphs. Then since Condition $2^+$ is stronger than Condition 2, we can tell that Stage 2 actually works for all of the 711,200 tested graphs.

By the definition of $N[uv]$ in Condition $2^+$, 3-cliques of $S$ in Step 4 and $SS$ in Step 5 require computation, thus the running time for the algorithm that reaches Stage 2 as its final stage is $O(n^{3(5+\log n)/2})$.

However, the definition of $N[uv]$ in Condition $2^+$ looks "strange", so it requires an expression in a more general way as shown in the following section.

## 6. FURTHER STAGES OF THE ALGORITHM

Because it has not been theoretically proved that Stage 2 of the algorithm is applicable to all graphs, it is worthwhile to conceive further stages of the algorithm which shall be in a general form that is consistent with Stages 1 and 2. Then the concept of hyperedge of hypergraph is needed.

A hypergraph consists of a collection of vertices and a collection of hyperedges. If the vertex set is $V$, then the hyperedges are subsets of $V$. Then for Stage $t$ of the algorithm, where $t$ is a positive integer, we give the following five rules for the algorithm:

1. Suppose $E_s(H)$ is the set of all hyperedges of size $s$ in $H$, then let
   $E_s(H) = \{e : e$ is the vertex set of an $s$-clique in $H$ and $H$ is generated by Steps 1 to 3$\}$, $\forall\, s \le t+1$;
   $E_s(H) = \varnothing$, $\forall\, s > t+1$.
2. For any hyperedge $e$ of size no larger than $t$, let $N[e] = \{x : \{x\} \cup e$ is a subset of a hyperedge of $H\}$.
3. For any hyperedge $e$ of size $t+1$, let $N[e] = \{x : x \in N[\{v\}], \forall\, v \in e\}$.
4. By deleting hyperedges in $H$, subgraph $S$ in Step 4 and subgraph $SS$ in Step 5 shall satisfy

   **Condition *X*.** For any hyperedge $e$ of the subgraph, there exists a vertex cover $C$ of $G$ such that $L(N[e]) = L(C)$.
5. If Stage $t$ does not yield clique at Step 5, start Stage $t+1$ at Step 4 for the component of $H$ where Stage $t$ ends.

Then it is not difficult to show that the above rules are consistent with Stages 1 and 2 of the algorithm.

Claims 2.8 and 2.10 can be easily proved to hold for any stage of the algorithm, so Stage $t$ yields a minimum vertex cover of graph $G$ at Step 6 if it yields a clique at Step 5.

Because in Stage $t$, hyperedges of size $t+1$, i.e., vertex sets of $(t+1)$-cliques, of $S$ in Step 4 and $SS$ in Step 5 require computation, the algorithm runs in $O(n^{(t+1)(5+\log n)/2})$ time if the "maximum" stage to which it reaches is Stage *t*.

## 7. CONCLUSIONS

Conclusions for this algorithm are summarized as follows.

*A.* For minimum vertex cover problem of an order-*n* graph, the algorithm runs in $O(n^{(t+1)(5+\log n)/2})$ time if the "maximum" stage to which it reaches is Stage *t*, where *t* is a positive integer. Therefore, if the "maximum" stage to which the algorithm reaches for a graph is Stage 1 or 2, then the algorithm runs in $O(n^{5+\log n})$ or $O(n^{3(5+\log n)/2})$ time for the graph respectively.

*B.* Stage 1 of the algorithm is applicable to, i.e., yields a proved minimum vertex cover for, each of more than 99% of tested graphs of order no larger than 32, while Stage 2 of the algorithm works for 100% of tested graphs of order no larger than 32. And it is reasonable to believe that the above ratios are close to the "real applicability ratios" for this algorithm to all unlabeled (or labelled) graphs of order no larger than 32.

*C.* The "applicability ratio" of Stage 1 for all tested order-8 graphs, order-16 graphs and order-32 graphs is 100%, 99.99% and 99.67% respectively, so it is unlikely that the "applicability ratio" of Stage 1 decreases sharply when the order of graph increases. Besides, Stage 2 works for each of the tested graphs and we still have stages beyond Stage 2. Therefore it is an efficient algorithm which is already applicable to practical use.

*D.* As summarized above, the performance of this algorithm is extraordinary for NP-Complete problems. Although there is no systematic and theoretical explanation for it yet, it is at least an important finding which is valuable for further research. However, like many findings or conjectures in mathematics, its theoretical explanation may take many years even decades to be found. Thus it is decided to make the algorithm public so that the explanation and improvement of it become possible.

*E.* Further tests for graphs of order 64 or more are necessary and valuable. However, they require computer of large capacity which is not available for the author.

## Appendix 1. Generating_k_and_H.m

```
% This MATLAB file is the first program to test the algorithm in the
manuspcript and shall run before the running of file "Testing_G".

function Generating_k_and_H
k=input('When the order of each randomly generated graph is 2^k, k equals
to:');
if (k<=0)||(k<ceil(k))
    error('k must be positive interger')
end
if k==6
    disp('Note: This program requires large memory when k is larger than
5. The size of the swap file is recommended to be set to 100GB-200GB for
k=6.')
```

```
    disp('And the running time for k=6 on an ordinary personal computer
is about 2 to 3 hours.')
end
if k>=7
    disp('Note: For k>6, this program requires large memory which may
exceed the capacity of the computer,')
    disp('and the running time on an ordinary personal computer is very
long.')
end
if k>7
    error('For k>7, all of the uint8 class of matrix in this program shall
be changed to other class.')
end
Z=Generating_Z(k);
H=Generating_H(k,Z);
disp('Saving file "k_and_H". Please wait.')  % When k is larger than 5,
the following saving of file 'k_and_H' may take a long time.
save('k_and_H','k','H','-v7.3')
end

% Step 1
function Z=Generating_Z(k)
% Cell-array Z represents the family of sets by Definition 2.1 of the
algorithm, while its non-empty cells are arrays which represent the members
of
% family Z. Z's cells of size 1 are first generated as follows.
Z=cell(k+1,2^k);
for c=1:2^k
    M=c;
    Z{1,c}=M;
end
% Z's cells of larger size are generated as follows. Non-empty cells of
the rth row of cell-array Z represent family Z's member of size 2^(r-1).
for r=2:k+1
    for c=1:2^(k+1-r)
        Z{r,c}=union(Z{r-1,2*c-1}, Z{r-1,2*c});
    end
end
end

% Steps 2&3
function H=Generating_H(k,Z)
% In this program, component P of graph H is expressed by its adjacency
matrix where diagonal numbers are the labels of corresponding vertices
```

```
% instead of zeros. For first-grade component P(Z{1,d2}), H{1,d2,1,1} is assigned to be its adjacency matrix. However, for higher-grade component
% P(Z{d1,d2},d3,d4) where d1>1, H{d1,d2,d3,d4} is assigned to be its adjacency matrix.
H=cell(k+1,2^k,2^(k-1),2^(k-1));
% k+1=size(Z,1). 2^k=size(Z,2).
% When component P's grade is larger than 2^(k-1), P is the join of same-grade unions U(Z{k,1},d3) and U(Z{k,2},d4). Because |Z{k,1}|=|Z{k,2}|=2^(k-1),
% so both d3 and d4 are no larger than 2^(k-1) for any H{d1,d2,d3,d4}.
U=cell(k+1,2^k,2^(k-1));
% Same-grade union U(Z{d1,d2},D3) is expressed by its adjacency matrix U{d1,d2,D3} where diagonal numbers are the labels of corresponding vertices
% instead of zeros.
% Matrices for the first-grade components and first-grade unions are first generated as follows.
for d2=1:2^k
    H{1,d2,1,1}=uint8(d2);  % The uint8 class is used to save memory and time. However, it shall be changed to other class when k is larger than 7.
    U=Generating_U(U,k,H{1,d2,1,1},1,d2,1);
end
% Matrices for higher-grade components and higher-grade unions are generated as follows.
for b=2:2^k
    disp(['Proceeding to b = ',num2str(b),' of ',num2str(2^k)])  % This command is to show the progress of the running program when k is large.
    for d3=max(1,b-2^(k-1)):min(b-1,2^(k-1))
        % b=d3+d4. b3 and b4 are positive integers and both maximum values of d3 and d4 for H{d1,d2,d3,d4} are 2^(k-1).
        d4=b-d3;
        for d1=1:k
            if (size(Z{d1,1},2)<d3)||(size(Z{d1,1},2)<d4)
                % By Definition 2.5, for component P(Z{d1+1,*},d3,d4) and nonempty H{d1+1,*,d3,d4}, d3 and d4 are not larger than size(Z{d1,1},2).
                continue
            end
            for d2=1:2:2^k-1
                if isempty(Z{d1,d2+1})==0
                    % Z{d1,d2} and Z{d1,d2+1} represent A1 and A2 in Definition 2.5, respectively.
                    sx=size(U{d1,d2,d3},1); % Matrix U{d1,d2,d3} expresses same-grade union U(Z{d1,d2},d3)
```

```
                 sy=size(U{d1,d2+1,d4},1); % Matrix U{d1,d2+1,d4}
expresses same-grade union U(Z{d1,d2+1},d4)

H{d1+1,(d2+1)/2,d3,d4}=uint8([U{d1,d2,d3},ones(sx,sy);ones(sy,sx),U{d
1,d2+1,d4}]);
                 % H{d1+1,(d2+1)/2,d3,d4} is the adjacency matrix of
graph H's component P(Z{d1+1,(d2+1)/2},d3,d4), which is the join of
                 % U(Z{d1,d2},d3) and U(Z{d1,d2+1},d4).
                 % Z{d1+1,(d2+1)/2}=union(Z{d1,d2},Z{d1,d2+1}).
                 if d1<k
                     % When d1=k, H{d1+1,*,*,*} expresses component of
same-grade union U(Z{k+1,1},*). Because |Z{k+1,1}| = 2^k, i.e., Z{k+1,1}
                     % contains all labels, so U(Z{k+1,1},*) will not be
used to generate any component of H and thus they are not generated here.

U=Generating_U(U,k,H{d1+1,(d2+1)/2,d3,d4},d1+1,(d2+1)/2,b);
                 end
             else % which indicates Z{d1,d2+1} is empty, then go to the
next d1.
                 break
             end
         end
     end
   end
end
end

function U=Generating_U(U,k,P,a1,a2,b)
% This function is to put component P(Z{a1,a2},b1,b2) into each same-grade
union U(Z{d1,d2},b) of which b=b1+b2 and Z{a1,a2} is a subset of Z{d1,d2}.
d1=a1;
d2=a2;
while d1<=k
    % As explained in lines 80 and 81, U(Z{k+1,*},*) will not be used to
generate any cell of H and thus they are not generated here.
    sizeU=size(U{d1,d2,b},1);
    sizeP=size(P,1);

U{d1,d2,b}=uint8([U{d1,d2,b},zeros(sizeU,sizeP);zeros(sizeP,sizeU),P]
);
    % The above command puts component P(Z{a1,a2},b1,b2) into same-grade
union U(Z{d1,d2},b).
    d1=d1+1;  % It is obvious that for any d1, there exists at most one
Z{d1,d2} which includes Z{a1,a2}.
```

```
    d2=ceil(d2/2); % If Z{d1,d2} includes Z{a1,a2}, then
Z{d1+1,ceil(d2/2)} includes Z{a1,a2}.
end
end
```

**Appendix 2. Testing_G.m**

```
% This MATLAB file is the second program to test the algorithm in the
manuspcript. Another file named "Generating_k_and_H.m" needs to run
first.

function Testing_G
% By generating m random graphs, of which the order is 2^k, the program
finds out how many of the tested random graphs do not yield clique in Step
5
% and thus are not the graphs to which the algorithm is applicable. (If
a clique is yielded in Step 5, then by Claim 2.10, its label set is the
label
% set of a minimum vertex cover of the tested graph.)
m=input('How many random graphs are to be generated and tested for the
algorithm?');
if (m<=0) || (m<ceil(m))
    error('m must be positive integer')
end
disp('Loading file "k_and_H". Please wait.')  % When k is larger than 5,
the following loading of file 'k_and_H' may take a long time.
load('k_and_H');   % File "k_and_H.mat" is generated by
"Generating_k_and_H.m".
if k==5
    disp('Note: When k=5, the running time for an ordinary personal
computer to compute a graph is about 9 minutes.')
end
if k>5
    disp('Note: When k>5, the running time for an ordinary personal
computer to compute a graph is very long.')
end
n1=0; % n1 will be the number of tested graphs to which Stage 1 of the
algorithm is not applicable.
n2=0; % n2 will be the number of tested graphs to which Stage 2 is not
applicable.
d1=[]; % d1 will record the approximate edge density of each graph to which
Stage 1 is not applicable.
d2=[]; % d2 will record the approximate edge density of each graph to which
Stage 2 is not applicable.
```

```
for h=1:m
    disp(['Computing the ',num2str(h),'th random graph'])
    density=2;
    while density<=0 || density>1
        density=0.5+0.3*randn;
    end
    % "density" will approximately equals to the ratio of |E(G)| to |E(K)|,
of which K is a 2^k-order complete graph, and it is generated from a normal
    %  distribution with mean 0.5 and standard deviation 0.2,0.3,0.4,0.5
or 0.6.
    G=Generating_G(k,density);
    if G==zeros(size(G))
        disp(['The ',num2str(h),'th random graph has no edge.'])
        continue
    end
    stage=1;
    bstart=2;
    [S1,bstart]=Step_4(stage,bstart,k,G,H);
    I1=Step_5(stage,S1,G);
    result1=Step_6(I1);
    if result1==1  % which indicates I1 is a clique.
        disp(['Stage 1 of the algorithm is applicable to the
',num2str(h),'th random graph.']);
    else
        disp(['Stage 1 of the algorithm is NOT applicable to the
',num2str(h),'th random graph.']);
        n1=n1+1;
        d1(length(d1)+1)=density;
        stage=2;
        [S2,~]=Step_4(stage,bstart,k,G,H);
        I2=Step_5(stage,S2,G);
        result2=Step_6(I2);
        if result2==1  % which indicates I2 is a clique.
            disp(['Stage 2 of the algorithm is applicable to the
',num2str(h),'th random graph.']);
        else
            disp(['Stage 2 of the algorithm is NOT applicable to the
',num2str(h),'th random graph.']);
            n2=n2+1;
            d2(length(d2)+1)=density;
        end
    end
    disp(['Currently, Stage 1 of the algorithm is NOT applicable to
',num2str(n1),' of ',num2str(h),' computed random graphs.'])
```

```
   disp(['Currently, Stage 2 of the algorithm is NOT applicable to
',num2str(n2),' of ',num2str(h),' computed random graphs.'])
end
d1
disp('d1 records the approximate edge density of each graph to which Stage
1 is not applicable.')
d2
disp('d2 records the approximate edge density of each graph to which Stage
2 is not applicable.')
end

function G=Generating_G(k,density)
G=sprandsym(2^k,density,0.1);
% sprandsym(n,density) is a symmetric random, n-by-n, sparse matrix with
approximately density*n*n nonzeros. However, when density equals to 1,
the
% function still generates some zeros. For better performance,
sprandsym(n,density,rc) is used here and it works much better, although
ocassionly still
% generates zeros, for density=1.
G(logical(eye(2^k)))=0;
G=spones(G);
end

function [S,bstart]=Step_4(stage,bstart,k,G,H)
% For each component P of graph H in ascending order of its grade, by
deleting a minimal subgraph of P, get P's maximal subgraph S which satisfies
% Conditions 1 and 2 (or 2+ for Stage 2). Iterate this computation for
the next component until P's S is nonempty.
% In this program, component P of graph H is expressed by its adjacency
matrix, where diagonal numbers are the labels of corresponding vertices
% instead of zeros. For first-grade component P(Z{1,d2}), H{1,d2,1,1} is
assigned to be its adjacency matrix. However, for higher-grade component
% P(Z{d1,d2},d3,d4) where d1>1, H{d1,d2,d3,d4} is assigned to be its
adjacency matrix.
if stage==1
   for d2=1:2^k  % This "for loop" is only for graph H's first-grade
components. Stage 2 never starts from any first-grade component because
a
       % nonempty subgraph of a first-grade component is always a clique.
       P=H{1,d2,1,1};

S=Maximal_subgraph_satisfying_conditions_1and2_or_1and2plus(stage,P,G
);
```

```
        if isempty(S)==0 % which indicates P(Z{1,d2}) has a nonempty
subgraph S, which must be P(Z{1,d2}) itself, that satisfies Conditions
1 and 2.
            return
        end
    end
end
for b=bstart:2^k  % This "for loop" is for graph H's components of grade
larger than 1. Thus bstart=2 for Stage 1, while Stage 2 starts at the grade
    % of H's component which has nonempty subgraph satisfying Conditions
1 and 2.
    for d3=max(1,b-2^(k-1)):min(b-1,2^(k-1))
        % b=d3+d4. d3 and d4 must be positive integers and both maximum
values of d3 and d4 for H{d1,d2,d3,d4} are 2^(k-1).
        d4=b-d3;
        for d1=2:k+1
            for d2=1:2^k
                P=H{d1,d2,d3,d4};
                if isempty(P)==0   % which indicates this cell represents
a nonempty component P(Z{d1,d2},d3,d4).

S=Maximal_subgraph_satisfying_conditions_1and2_or_1and2plus(stage,P,G
);
                    if isempty(S)==0  % which indicates P(Z{d1,d2},d3,d4)
has a nonempty subgraph S which satisfies conditions 1 & 2(or 2+ for Stage
2).
                        bstart=b;  % This command records the grade of H's
component at which the possible Stage 2 will start.
                        return
                    end
                end
            end
        end
    end
end
end

function I=Step_5(stage,S,G)
% For the recorded S of Step 4, by deleting a maximal subset of V(S), get
a minimal-order subgraph I of S which contains a nonempty subgraph SS that
% satisfies Conditions 1 and 2(or 2+ for Stage 2). Record this I for the
next step.
v=1;
while v<=size(S,1)
```

```
    % The vth vertex of S is to be checked whether it can be deleted, i.e.,
whether S still has a nonempty subgraph which satisfies Conditions 1 and
2
    % (or 2+ for Stage 2) after the vth vertex of S is deleted.
    I=S;
    I(v,:)=[];
    I(:,v)=[];

SS=Maximal_subgraph_satisfying_conditions_1and2_or_1and2plus(stage,I,
G);
    if isempty(SS)==0  % which indicates that I still has a nonempty
subgraph which satisfies Conditions 1 and 2(or 2+ for Stage 2).
        S=SS;  % If this command were S=I instead of S=SS, then the next
round of function
"Maximal_subgraph_satisfying_conditions_1and2_or_1and2plus"
        % would delete again the edges of E(I)-E(SS).
        % For any i which satisfies 1 <= i < v, it is obvious that the ith
vertex of the last S remains not deletable and so it is still the ith vertex
        % of the current S.
    else  % which indicates the vth vertex of S shall not be deleted.
        v=v+1;
    end
end
I=S;
end

function result=Step_6(I)
% This function checks whether I represents a clique.
if size(I,1)==1
    result=1;
    return
end
for i=1:size(I,1)-1  % Because of symmetry of matrix I, only the "upper
half" of the matrix is checked.
    for j=i+1: size(I,1)
        if I(i,j)==0
            result=0;
            return;
        end
    end
end
result=1;
end
```

```
function
M=Maximal_subgraph_satisfying_conditions_1and2_or_1and2plus(stage,D,G
)
% For the graph defined by adjacency matrix D, this function finds out
its maximal subgraph which satisfies Conditions 1 and 2 (or 2+ for Stage
2).
if isempty(D)==1
    M=D;
    return
end
x=1;
while x==1
    DD=A_round_of_edge_deleting_by_Condition_2or2plus(stage,D,G);
    if DD==D % which indicates all of the edges of the graph expressed
by D satisfy Condition 2 (or 2+ for Stage 2).
        break
    end
    % At this point, DD~=D, which indicates certain edges of the graph
expressed by D were deleted in the last round of edge-deleting and D becomes
DD.
    % However,the graph expressed by DD may still have edges which do not
satisfy Condition 2 (or 2+ for Stage 2), so the next round of edge-deleting
    % is necessary until DD=D.
    D=DD;
end
% At this point, all edges of the graph expressed by D satisfy Condition
2 (or 2+ for Stage 2). Then it is obvious that for any vertex which has
% incident edge, Condition 1 is satisfied. Thus the following part of this
function is to delete the isolated vertices which do not satisfy Condition
1.
i=1;
while i<=size(D,1)  % The ith vertex of D will be checked whether it is
an islated vertex and whether it satisfies Condition 1.
    A=D(i,:); % Thus A(i)=D(i,i) is the label of the ith vertex of D.
    A(i)=0;
    if A==zeros(size(A))
        % which indicates the ith vertex of D has no incident edge, then
we need to check whether the label of this vertex is also the label of
a G's
        % vertex cover of size 1 as follows.
        GG=G;
        GG(D(i,i),:)=0;  % The label of the xth vertex of G is x.
        GG(:,D(i,i))=0;
        % D(i,i) is the label of the ith vertex of D. Suppose j=D(i,i).
```

```
If j is the label of a G's vertex cover of size 1, then for any G(r,c)~=0,
       % r=j or c=j. (Note that all of the diagonal numbers of matrix G
are zeros.)
       if GG==zeros(size(G))
          % which indicates the label of the ith vertex of D is the label
of a G's vertex cover of size 1, then this vertex satisfies Condition 1
          % and shall not be deleted.
          i=i+1;
       else  % Then this vertex does not satisfy Condition 1 and shall
be deleted.
          D(i,:)=[];
          D(:,i)=[];
       end
   else  % which indicates the ith vertex of D has incident edge and shall
not be deleted.
       i=i+1;
   end
end
M=D;
end

function DD=A_round_of_edge_deleting_by_Condition_2or2plus(stage,D,G)
% For each round of edge-deleting, all edges which do not satisfy Condition
2 (or 2+ for Stage 2) of the input graph expressed by D are deleted.
% However, the deletion may cause some more edges which no longer satisfy
Condition 2 (or 2+ for Stage 2), and these edges may not be all deleted
in
% this round of edge-deleting.
i=2;
while i<=size(D,1)
   for j=1:i-1  % Because matrix D is symmetrical, only the part under
its diagonal is checked.
       if D(i,j)==1
          dia=diag(D); % dia(1,x) is the label of the xth vertex of D.
          Ne2=D(i,:).*D(j,:);
          % ".*" is element-by-element multiplication, so Ne2(1,x) is
non-zero only if both D(i,x) and D(j,x) are non-zero. Therefore, Ne2(1,x)
is
          % non-zero only if x=i, x=j, or the xth vertex is adjacent to
both the ith and the jth vertices, so Ne2=N[ij] for Stage 1.
          if stage==2  % then Condition 2+ shall be satisfied instead of
Condition 2.
              for k=1:size(D,1)
                 if k~=i && k~=j && Ne2(1,k)==1  % which indicates the
```

```
kth vertex is adjacent to both the ith and the jth vertices.
                        Ne3=Ne2.*D(k,:); % Similar as explained for Ne2,
Ne3(1,x) is non-zero only if x=i, x=j, x=k, or the xth vertex is adjacent
to
                        % all of the ith, jth and kth vertices.
                        LNe3=dia(logical(Ne3)); % LNe3 is the label set of
Ne3.
                        C=Does_the_label_set_cover_E_of_G(LNe3,G);
                        if C==0  % which indicates the 3-clique of vertices
i,j and k does not satisfy the definition of N[uv] in Condition 2+, so
                            % vertex k shall not be considered to be an element
of N[ij] for Stage 2.
                            Ne2(1,k)=0;
                        end
                    end
                end
                % At this point, Ne2=N[ij] for Stage 2.
            end
            LNe2=dia(logical(Ne2));
            % LNe2 is the label set of Ne2=N[ij].
            C=Does_the_label_set_cover_E_of_G(LNe2,G);
            if C==0   % which indicates the edge corresponding to D(i,j)
does not satisfy Condition 2 (or 2+ for Stage 2).
                D(i,j)=0;
                D(j,i)=0;
            end
        end
    end
    i=i+1;
end
DD=D;
end

function C=Does_the_label_set_cover_E_of_G(LN,G)
% This function checks whether LN is a label set of a vertex cover of G.
G(LN,:)=0;
G(:,LN)=0;
if G==zeros(size(G))
    C=1;
else
    C=0;
end
end
```